\documentstyle[12pt,epsfig]{article}
\begin{document}

\thispagestyle{empty}

\rightline{FNT/T-99/05} \vskip 24pt

\begin{center}  {\Large\bf   On Large-Angle Bhabha Scattering at LEP}
\end{center}

\noindent \begin{center} Guido MONTAGNA$^{a,b}$, Oreste NICROSINI$^{b,a}$     
and Fulvio PICCININI$^{b,a}$  \end{center}

\noindent \begin{center} $^a$ Dipartimento di Fisica Nucleare e Teorica, 
Universit\`a di Pavia, \\
Via A.~Bassi 6, 27100, Pavia, Italy \\ 
$^b$ INFN, Sezione di Pavia,  Via A.~Bassi 6, 
27100, Pavia, Italy  
\end{center}

\vskip 64pt

\begin{abstract}{\small\noindent  The theoretical accuracy of the program
TOPAZ0 in the large-angle Bhabha channel is estimated. The physical error
associated with the full Bhabha cross section and its forward and backward
components separately is given for some event selections and several energy
points of interest for LEP1 physics, both for the $s$ and non-$s$
contributions to the cross section.   } \end{abstract}

\vskip 64pt \noindent E-mail: \\ montagna@pv.infn.it \\ nicrosini@pv.infn.it
\\ piccinini@pv.infn.it \\

\vfill \leftline{FNT/T-99/05} \leftline{April 19, 1999}

\newpage\normalsize

One of the open issues of precision physics at LEP 
is the determination of the
accuracy of the theoretical predictions for the large-angle Bhabha scattering
cross section. At present, several computer codes developed for large-angle
Bhabha scattering studies  can be found in the literature, ranging from 
semi-analytical to truly Monte Carlo ones. A detailed account 
of them has been
presented in refs.~\cite{lep2bha,rassegna}. In particular, in 
ref.~\cite{lep2bha}   several comparisons have been
performed, both for ``academic'' and realistic event selections (ES's), 
both for LEP1 and LEP2 energies.

After the publication of ref.~\cite{lep2bha}, a new analysis concerning
specifically two codes, namely ALIBABA~\cite{alibaba} and
TOPAZ0~\cite{topaz0}, has been performed~\cite{bp98}, where a very detailed 
comparison between the two programs is developed and the estimate of the
theoretical error associated with their predictions is given. 
If, on the one hand,
the comparison is very careful, on the other hand the study 
is, in the opinion
of the authors of the present note, lacking for the following aspects: it
considers, as the  main source of information on large-angle Bhabha
observables, only ALIBABA and TOPAZ0; it is based on a comparison for a
``bare'' ES, which is far from being realistic, and ignores more realistic
ES's such as the ones considered  in ref.~\cite{lep2bha};  
it does  not fully exploit the
detailed comparisons for $s$-channel annihilation 
processes, that can be found
in the literature~\cite{9503} and give  valuable pieces of information on a
significant part of the full Bhabha cross section;  
it considers only centre
of mass energies around the $Z^0$ resonance, leaving 
aside LEP2 energies, from
which additional information can be extracted concerning the accuracy
of the non-$s$ component of the full Bhabha cross section; 
it addresses the
problem of assigning a theoretical error to the full Bhabha 
cross section both
for ALIBABA and TOPAZ0, but the error for the  non-$s$   part of the Bhabha
cross section is given for ALIBABA only;  moreover no information is given
concerning the forward and  backward components of the cross section
itself.    

\begin{table}[hbt]

\begin{tabular}{|l||c|c|c|c|c|c|c|c|}

\hline $\sqrt{s} $ (GeV) & 88.45 & 89.45 & 90.20 & 91.19 & 91.30 & 91.95 &
93.00  & 93.70 \\ \hline \hline $\sigma^T$ (pb)& 457.08 & 644.86 & 912.06 &
1185.70 & 1164.82 & 873.50  & 476.64& 351.80\\ \hline $\sigma_s^T$ (pb)&
172.94 & 331.55 & 590.93 & 994.27 & 998.32 & 820.80  & 461.49 & 329.61 \\
\hline  $\sigma_{ns}^T$ (pb)& 284.14 & 313.31 & 321.13 & 191.43 & 166.50 &
52.70  & 15.15 & 22.19 \\ \hline  $\sigma_{ns}^A$ (pb)& 284.11 & 312.69 &
320.71 & 192.66 & 166.93 & 55.27  & 16.88 & 23.82 \\ \hline 
$\delta\sigma_s^T$ (pb)&  0.2 & 0.3 & 0.6 & 1.0 & 1.0  & 0.8 & 0.5 &  0.3 \\
\hline $\delta\sigma_{ns}^T$ (pb)& 2.8  &  3.1 & 3.2 & 1.9  & 1.7 & 2.6 &
1.7   &  1.6 \\ \hline $\delta\sigma / \sigma$ (\%)& \bf 0.7 & \bf 0.5 & \bf
0.4 & \bf  0.2  & \bf 0.2 & \bf 0.4 & \bf 0.5 & \bf  0.5\\ \hline
\end{tabular} 
\caption{Estimate of the theoretical error of TOPAZ0 for the full 
Bhabha cross 
section at $10^\circ$ maximum acollinearity. $\sigma^T$, $\sigma_s^T$  and 
$\sigma_{ns}^T$ are the full, $s$ and non-$s$ part  of the  TOPAZ0 cross
section. $\sigma_{ns}^A$ is the non-$s$ part  of the  ALIBABA cross
section.  $\delta\sigma_s^T$ and $\delta\sigma_{ns}^T$ are the absolute 
theoretical error of the  $s$ and non-$s$  parts of the 
TOPAZ0 cross section,
as obtained  according  to the procedure given in  the text.  
$\delta\sigma / \sigma$  is   the  total  relative error. } 
\label{tab:stbha10} 
\end{table}

The aim of the present study is to critically analyze as much as possible of
the available literature on large-angle Bhabha scattering, in order to give a
reliable estimate of the theoretical error associated with TOPAZ0, both for
full cross sections and for the forward and backward components, both for
the $s$ and non-$s$ parts.  The numerical results presented in the following 
are obtained mostly by elaborating on the ALIBABA and TOPAZ0 predictions  
shown in ref.~\cite{bp98}.      

\begin{table}

\begin{tabular}{|l||c|c|c|c|c|c|c|c|}

\hline $\sqrt{s} $ (GeV) & 88.45 & 89.45 & 90.20 & 91.19 & 91.30 & 91.95 &
93.00  & 93.70 \\ \hline \hline $\sigma^T$ (pb)& 485.17 & 674.89 & 945.00 &
1221.13 & 1200.16 & 905.25  & 503.79& 377.59\\ \hline $\sigma_s^T$ (pb)&
176.31 & 336.84 & 599.25 & 1007.03 & 1011.10 & 831.43  & 468.16 & 334.94 \\
\hline  $\sigma_{ns}^T$ (pb)& 308.86 & 338.05 & 345.75 & 214.10 & 189.06 &
73.82  & 35.63 &  42.65\\ \hline  $\sigma_{ns}^A$ (pb)& 306.62 & 335.47 &
343.44 & 212.36 & 187.84 & 73.86  & 35.16 &  42.48\\ \hline 
$\delta\sigma_s^T$ (pb)&  0.2 & 0.3 & 0.6 & 1.0 & 1.0  & 0.8 & 0.5 &  0.3 \\
\hline $\delta\sigma_{ns}^T$ (pb)&  3.1 & 3.4  & 3.5 &  2.1 & 1.9 & 0.7 & 0.5
& 0.4 \\ \hline $\delta\sigma / \sigma$ & \bf 0.7 & \bf  0.5& \bf 0.4 & \bf 
0.3& \bf 0.2  & \bf 0.2 & \bf 0.2 & \bf 0.2 \\ \hline 
\end{tabular}
\caption{The same as tab.~\ref{tab:stbha10} at  $25^\circ$ maximum 
acollinearity. }
\label{tab:stbha25} 
\end{table}

Let us consider first the problem of assigning a theoretical  error to the
full  $s+t$ Bhabha cross section.  It can be decomposed into $s$ and non-$s$ 
contributions.  As far as the $s$ part is concerned, TOPAZ0 includes exact
$O(\alpha)$ electroweak  corrections  plus all the relevant and 
presently under control higher order 
contributions. From several tuned comparisons discussed in recent
literature~\cite{9503}, 
one  can see that the overall difference between TOPAZ0 and
ZFITTER~\cite{zfitter}   is at the scale of $0.01\%$ for $s$ channel QED
convoluted cross sections, for both extrapolated and realistic set up. 
While for completely inclusive $s$-channel  cross sections, or for 
$s$-channel  cross sections with an $s'$ cut, TOPAZ0 includes  
$O(\alpha^3 L^3)$  and $O(\alpha^2 L)$ hard photon  corrections according to 
ref.~\cite{a3a2l}, 
these are not taken into account for $s$-channel  processes with angular 
acceptance cuts and, in  particular, for the $s$ part of the full Bhabha 
cross section. When taking into account  the theoretical error due
to neglecting them, and  the one due to other minor sources such as  
the approximate treatment  of 
additional  light pairs, one can
conclude that the overall theoretical error of the 
$s$ part of the Bhabha cross
section in TOPAZ0 is $0.1\%$. In setting the theoretical error for the $s$
part, no information coming from ALIBABA is considered, because  it is  
known that the code is not accurate for $s$-channel processes 
at the 0.1\% level~\cite{lep2bha,topaz0} as, for instance, 
TOPAZ0 and ZFITTER
are.  As far as the non-$s$
part is  concerned, the theoretical error of TOPAZ0 is dominated by missing
$O(\alpha)$  non-logarithmic QED corrections, that on 
the contrary are present
in ALIBABA.   A way of estimating such an error is 
to consider the comparisons
between TOPAZ0 and  BHWIDE~\cite{bhwide} performed in 
ref.~\cite{lep2bha} at LEP2
energies. Actually, in the  LEP2 energy regime Bhabha scattering is
essentially a $t$-channel dominated  process. Since  BHWIDE contains exact
$O(\alpha)$ QED corrections for the $s$ and non-$s$ 
contributions to the cross
section, such a comparison sets the size of the  missing
non-log contributions in the non-$s$ part of the TOPAZ0 cross section, 
which is at the 1\% level. In
order to be as much as  conservative as possible, and not to  loose the
information contained in ALIBABA for the non-$s$ contributions,  a reliable
recipe for setting the  theoretical error of TOPAZ0 for the non-$s$ part of
the Bhabha cross section is to  take  it as the maximum between 1\% of the
non-$s$ part of the cross section and the absolute deviation from ALIBABA. 

\begin{center}
\begin{table}[hbt]

\begin{tabular}{|l||c|c|c|c|c|c|c|c|}

\hline 
$\sqrt{s} $ (GeV) & 88.45 & 89.45 & 90.20 & 91.19 & 91.30 & 91.95  &
93.00 & 93.70 \\ 
\hline \hline 
$\sigma_F^T$ (pb)& 67.43 & 142.90 & 272.98 &
497.33  & 503.20 & 430.39 & 253.36 &184.99 \\ 
\hline 
$\sigma_B^T$  (pb)& 105.51
& 188.65 & 317.95 & 496.94  & 495.12 & 390.41 & 208.13 & 144.62 \\ 
\hline 
$\delta\sigma_F^T$ (pb)& \bf 0.1 & \bf 0.1 & \bf 0.3  & \bf 0.5 & \bf 0.5 &
\bf 0.4 & \bf 0.3 & \bf 0.2  \\ 
\hline 
$\delta\sigma_B^T$ (pb)& \bf 0.1 & \bf
0.2 & \bf 0.3  & \bf 0.5 & \bf 0.5 & \bf 0.4 & \bf 0.2 & \bf 0.1 \\ 
\hline
\end{tabular} 
\caption{Estimate of the absolute theoretical error of TOPAZ0 for the $s$ 
part of  the Bhabha cross 
section at $10^\circ$ maximum acollinearity, for the forward ($\sigma_F^T$) 
and  backward ($\sigma_B^T$) components separately. } 
\label{tab:sbha10} 
\end{table}
\end{center}

By following the above recipe for the estimate of  the theoretical error of
TOPAZ0 for the full Bhabha cross section, tables~\ref{tab:stbha10} and
\ref{tab:stbha25}  follow. As already stated, the numerical results shown
are   elaborated from ref.~\cite{bp98}, 
where all the details of the ES and input
parameters adopted can be found. 

The estimate of the theoretical error for the F+B cross section given in 
tabs.~\ref{tab:stbha10} and \ref{tab:stbha25} refers to a BARE ES.  
Anyway, it is worth considering  also the
results of the comparison between  BHWIDE and TOPAZ0 shown  
in ref.~\cite{lep2bha} for the  LEP1 energy range,
for both BARE and CALO ES's.  Actually, it  is known that ALIBABA does not
contain the bulk of the $O(\alpha^2 L)$ corrections, 
while BHWIDE and TOPAZ0 do,
by virtue of their factorized 
formulation~\cite{topaz0,bp98,a3a2l,bhwide}. Such
missing  corrections are, for instance, responsible 
of part of the theoretical  
error of  ALIBABA, that  above the $Z$ peak for $s$ channel processes can be 
of the order  of  several 0.1\%. When considering the comparison between
BHWIDE and TOPAZ0 for LEP1 energies, one  realizes 
that the  difference between
the two programs for a  realistic CALO ES is generally 
smaller than the errors
quoted above\footnote{The authors of BHWIDE consider the program  as more
reliable for realistic ES's (CALO) rather   
than for BARE  ones~\cite{bhwide}.}.  
Hence, the  estimate of the theoretical error of 
tabs.~\ref{tab:stbha10}  and \ref{tab:stbha25} has to  be considered as a
conservative one,  and has its origin in the fact that the 
recipe adopted aims at  using as  much information as  possible, and in 
particular the piece of information given by ALIBABA. 
In the light of the above
comments,  the error estimate 
 of tabs.~\ref{tab:stbha10}  and \ref{tab:stbha25}
can be considered as a  conservative error  estimate also for CALO ES's. 
Less conservative error estimates for  the full Bhabha cross section can be
found in  refs.~\cite{rassegna} and \cite{bhwide}.  

\begin{table}

\begin{tabular}{|l||c|c|c|c|c|c|c|c|}

\hline $\sqrt{s} $ (GeV) & 88.45 & 89.45 & 90.20 & 91.19  & 91.30 & 91.95 &
93.00 & 93.70 \\ \hline \hline $\sigma_F^T$ (pb)&  68.45 & 144.77 & 276.34 &
503.31  & 509.24 & 435.67 & 256.72 & 187.62\\ \hline $\sigma_B^T$ (pb)& 
107.86 & 192.07 & 322.91 & 503.72  & 501.86 & 395.76 & 211.44 & 147.32\\
\hline $\delta\sigma_F^T$ (pb)&  \bf 0.1 & \bf 0.1  & \bf 0.3 & \bf 0.5 &
\bf 0.5 & \bf 0.4 & \bf 0.3 & \bf 0.2 \\ \hline $\delta\sigma_B^T$ (pb)& 
\bf 0.1 & \bf 0.2  & \bf 0.3 & \bf 0.5 & \bf 0.5 & \bf 0.4 & \bf 0.2 & \bf
0.2\\ 
\hline 
\end{tabular} 
\caption{The same as   tab.~\ref{tab:sbha10} at $25^\circ$ maximum 
acollinearity. } 
\label{tab:sbha25} 
\end{table}

Besides a reliable estimate of the theoretical error  for the full $s+t$
Bhabha cross section,  it is also interesting to give the forward (F) and
backward (B)   parts of the cross section, together with their theoretical
error,  for  both the $s$ and non-$s$ components. 
For the program TOPAZ0, the
first part of the  task can be accomplished by  solving the system  
\begin{eqnarray}  
&&\sigma = \sigma_F + \sigma_B , \nonumber \\  
&&\sigma A_{FB}=\sigma_F - \sigma_B ,  
\end{eqnarray}   
where  $\sigma$ and $A_{FB}$ are the  cross section and  the
forward-backward asymmetry, respectively,  even if TOPAZ0 has  been designed
for computing cross sections and asymmetries directly. For the second part, 
{\it i.e. } assigning a theoretical error to the F/B components, one should 
notice that a naive error propagation can lead to artificially
overestimated errors.  Hence, in the following an alternative procedure is
proposed. 

\begin{table}

\begin{tabular}{|l||c|c|c|c|c|c|c|c|}

\hline $\sqrt{s} $ (GeV) & 88.45 & 89.45 & 90.20  & 91.19 & 91.30 & 91.95 &
93.00 & 93.70 \\ \hline \hline $\sigma_F^T$ (pb)&  262.77 & 289.51 & 296.69 &
177.69   & 154.80 & 50.39  & 15.90  & 22.36\\ \hline $\sigma_B^T$ (pb)& 21.37 
& 23.80  & 24.44 &  13.74   & 11.70 &  2.31  &  -0.75 &  -0.17 \\ \hline
$\sigma_F^{A1}$ (pb)&  263.16 & 289.26 & 296.64 & 179.00   & 155.58 & 52.42 & 
17.38 &  23.90\\ \hline $\sigma_B^{A1}$ (pb)& 20.95 &  23.43 &  24.07  &
13.66  &  11.35  & 2.85  &  -0.50  & -0.08 \\ \hline $\sigma_F^{A2}$ (pb)&
263.54  & 289.95 & 296.74 & 179.00   & 156.06 & 52.79 &  17.91 & 23.88 \\
\hline $\sigma_B^{A2}$ (pb)& 21.06 &  23.26 &  23.95   & 14.76 & 11.85   &
2.24  &  -0.27  &  -0.05 \\ \hline $\Delta_1\sigma_F$ (pb)& 0.39 &  0.25  & 
0.05   &  1.31  &  0.78 &  2.03  & 1.48 &  1.54  \\ \hline $\Delta_1\sigma_B$
(pb)&  0.42  &  0.37  &  0.37   &  0.08  &  0.35  &  0.54 &  0.25  & 0.09\\
\hline $\Delta_2\sigma_F$ (pb)& 0.77 &  0.44  &  0.05   &  1.31  & 1.26  &
2.40   & 2.01 &  1.52  \\ \hline $\Delta_2\sigma_B$ (pb)&  0.31  &  0.54  & 
0.49   &  1.02  &  0.15  &  0.07 &  0.48  & 0.12\\ \hline $\delta\sigma_F^T$
(pb)& \bf 2.6  &   \bf 2.9   &   \bf 3.0   &  \bf 1.8  &   \bf 1.5  &   \bf 
2.4   &  \bf 2.0  &  \bf  1.5  \\ \hline $\delta\sigma_B^T$ (pb)& \bf 0.4  &
\bf  0.5   &   \bf 0.5  & \bf   1.0   & \bf  0.4   &  \bf  0.5   &  \bf 0.5 
&  \bf 0.1 \\ \hline 
\end{tabular} 
\caption{Estimate of the absolute theoretical error of TOPAZ0 for the non-$s$ 
part of  the Bhabha cross 
section at $10^\circ$ maximum acollinearity, for the forward ($\sigma_F^T$) 
and  backward ($\sigma_B^T$) components separately. The apices $A1$ and $A2$
refer to the results of ALIBABA according to the procedures described in the
text. $\Delta $ are the absolute differences between  ALIBABA and TOPAZ0. } 
\label{tab:nsbha10}
\end{table}

First of all, by using the $s$ components of  cross section and asymmetry as
quoted in ref.~\cite{bp98} one can 
compute the $s$ components of  the forward and
backward cross sections. Since in general the $s$ component  of the F/B
asymmetry is small,  one can attribute to the F/B components of the 
$s$-channel cross 
section the same theoretical error as the one attributed to the  integrated
$s$-channel cross  section, namely 0.1\%. From this recipe, 
tabs.~\ref{tab:sbha10} and 
\ref{tab:sbha25}  follow.  

For the non-$s$ component of the cross
section, it should be  still desirable to 
exploit the information provided by
ALIBABA. To this aim, it has to be  noticed that two procedures can be
followed. The first one consists in solving the system  above as done 
for TOPAZ0
(ALIBABA1). The second one consists in computing  directly  
the F/B components
of the cross section, both for the full and $s$  parts (ALIBABA2). For the
first procedure, the results  of ref.~\cite{bp98} have been used. 
For the second 
one, ALIBABA has been re-run with high numerical 
precision in order to neglect
the  integration error. The theoretical error 
to be attributed to the non-$s$
part of the  F/B cross section, similarly to what has been done for the F+B
cross section,  has  then been defined as the maximum among the 1\% of the
corresponding cross section and the absolute deviation  from ALIBABA1 and
ALIBABA2. By following the recipe here described, 
tabs.~\ref{tab:nsbha10} and \ref{tab:nsbha25} follow. 

\begin{table}

\begin{tabular}{|l||c|c|c|c|c|c|c|c|}

\hline $\sqrt{s} $ (GeV) & 88.45 & 89.45 & 90.20  & 91.19 & 91.30 & 91.95 &
93.00 & 93.70 \\ \hline \hline $\sigma_F^T$ (pb)&  285.84 & 312.6  & 319.62 &
198.9  & 175.87 & 70.14  & 35.08 &  41.50\\ \hline $\sigma_B^T$ (pb)& 23.02 & 
25.45 &  26.13  &  15.20 &  13.19 &  3.68  &  0.55  &  1.15 \\ \hline
$\sigma_F^{A1}$ (pb)&  283.82 & 310.17 & 317.71  & 197.09 & 174.52 & 70.09  &
34.54 &  41.22 \\ \hline $\sigma_B^{A1}$ (pb)&  22.80 &  25.30  & 25.73  & 
15.27  & 13.32  & 3.77   & 0.62  &  1.26  \\ \hline $\sigma_F^{A2}$ (pb)&
284.18  & 310.87 & 317.24  & 197.64 & 173.84 & 69.91  & 34.80 & 40.86  \\
\hline $\sigma_B^{A2}$ (pb)&  22.67 &  24.98  & 25.59  &  16.14  & 13.20  & 
3.43  &  0.94 &  1.20  \\ \hline $\Delta_1\sigma_F$ (pb)&  2.02  &  2.43  & 
1.91   &  1.81  &  1.35  &  0.05  &  0.54  &  0.28 \\ \hline
$\Delta_1\sigma_B$ (pb)& 0.22  &  0.15  &  0.40   &  0.07  & 0.13  & 0.09 &
0.07  & 0.11 \\ \hline $\Delta_2\sigma_F$ (pb)& 1.66   &  1.73  & 2.38    & 
1.26  &  2.03  &  0.23  &  0.28  & 0.64  \\ \hline $\Delta_2\sigma_B$ (pb)&
0.35  &  0.47  &  0.54   &  0.94  &  0.01 &  0.25 & 0.39  & 0.05 \\ \hline
$\delta\sigma_F^T$ (pb)& \bf 2.9  &   \bf 3.1   &   \bf 3.2  &  \bf  2.0  & 
\bf 2.0  &  \bf  0.7   &   \bf 0.5  &  \bf  0.6 \\ \hline $\delta\sigma_B^T$
(pb)& \bf 0.4  &   \bf 0.5    & \bf  0.5   &  \bf 0.9   &  \bf 0.1   & \bf 
0.3   &   \bf 0.4   & \bf  0.1\\ \hline 
\end{tabular} 
\caption{The same as tab.~\ref{tab:nsbha10} at  $25^\circ$ maximum
acollinearity. }
\label{tab:nsbha25} 
\end{table}

Two technical comments   are in 
order here.  The first one is that the recipe  adopted for the estimate  of
the non-$s$ theoretical error is sensible, since in some energy points  the
error  is  fixed to be  1\% of the corresponding cross section (typically in
the  region below  the $Z$ peak),  whereas in other ones 
it is fixed by one of
the absolute  differences (typically in the region at and above the $Z$ peak,
where the F/B components are numerically small).   The  second  comment
concerns the fact that  in reconstructing the  full  theoretical error of 
the  F+B cross section by adding its F/B components  
from tabs.~\ref{tab:sbha10}--\ref{tab:nsbha25}, for  the $s$  and non-$s$
parts, one obtains values that are equal or slightly  larger than the ones
obtained directly in tabs.~\ref{tab:stbha10} and  \ref{tab:stbha25}, as
expected. Of course,  the same remarks concerning 
the conservativeness of the 
error estimate for the F+B cross section apply to the F/B components
separately,  also. 

The present results
correspond to  an  angular acceptance of $40^\circ$--$140^\circ$ for  the
scattered electron. Since the sharing  between  the  $s$  and non-$s$ 
component  of the cross section depends
on the angular acceptance, the above estimate of the theoretical  error
can be considered valid for the presently adopted angular cuts; anyway, an  
important change in the angular cuts would require a reanalysis of the
situation, bearing in mind that for larger/narrower angular acceptances the
total error  can be expected to increase/decrease, respectively. 

To summarize, the estimate of the  theoretical error  of the Bhabha cross
section derived in the present letter is based upon the following pieces of
information: tuned comparisons between  TOPAZ0 and ZFITTER for  $s$-channel
observables~\cite{9503};  estimate  of  missing higher-order QED
corrections~\cite{a3a2l}; comparisons  between BHWIDE and TOPAZ0 for the full
Bhabha  cross section in the LEP1 and LEP2 energy  range~\cite{lep2bha}; 
comparisons
between ALIBABA  and TOPAZ0 for the non-$s$ part of the cross section.  The 
present paper updates  the existing literature for the theoretical error of
the  full F+B Bhabha cross section~\cite{bp98}, and improves it by providing
additional information on the  theoretical uncertainty to be associated  to
the F/B components of the  $s$/non-$s$  parts of the Bhabha cross section. 

\vskip 12pt  \noindent 
{\bf Acknowledgements} The authors are  indebted with
Marta  Calvi and  Giam\-pie\-ro Passarino for stimulating discussions on the 
subject.  



\begin{thebibliography}{9}

\bibitem{lep2bha} {S. Jadach, O.~Nicrosini et al., ``Event Generators for
Bhabha  Scattering'', in {\it Physics at LEP2},  G.~Altarelli, T.~Sj\"ostrand
and F.~Zwirner, eds.,  CERN Report {\bf 96-01} (Geneva, 1996), 
vol.~2, p.~229.}

\bibitem{rassegna} {G.~Montagna, O.~Nicrosini and F.~Piccinini, ``Precision 
Physics at   LEP'', Rivista del Nuovo Cimento {\bf 21} n.~9 (1998) 1; {\tt
hep-ph/9802302}. }

\bibitem{alibaba} {W.~Beenakker, F.~Berends and S.C. van der Mark,
Nucl.~Phys.  {\bf B349} (1991) 323. }

\bibitem{topaz0} {G.~Montagna, O.~Nicrosini,  G.~Passarino F.~Piccinini and
R.~Pittau,
Comput.~Phys.~Commun.~{\bf 76} (1993) 328,  
Nucl.~Phys.~{\bf B401} (1993 ) 3; \\
G.~Montagna, O.~Nicrosini,  G.~Passarino and F.~Piccinini, TOPAZ0 
version 2.0, 
in Comput.~Phys.~Commun.~{\bf 93} (1996) 120; \\ 
TOPAZ0 version 4.0, in
Comput.~Phys.~Commun.~{\bf 117} (1999) 278. }

\bibitem{bp98} {W.~Beenakker and G.~Passarino, Phys.~Lett.~{\bf B425} (1998)
199.} 

\bibitem{9503}{D.~Bardin, M.~Gr\"unewald and G.~Passarino, {\it Precision 
Calculation Project Report},  {\tt hep-ph/9902452};  \\
D.~Bardin et al., ``Electroweak Working Group Report'',  in
{\it Precision Calculations for the $Z$ resonance}, D.~Bardin, 
W.~Hollik and
G.~Passarino, eds., CERN Report  {\bf 95-03} (Geneva, 1995), p.~7, {\tt
hep-ph/9709229}. }

\bibitem{zfitter} {D.~Bardin et al., Z.~Phys. {\bf C44} (1989) 493; Comput.
Phys. Commun. {\bf 59} (1990) 303; Nucl.~Phys. {\bf B351} (1991) 1;
Phys.~Lett. {\bf B255} (1991)  290; CERN-TH~6443/92. }

\bibitem{a3a2l}{G.~Montagna, O.~Nicrosini and F.~Piccinini,  
Phys.~Lett.~{\bf B406} (1997) 243; Phys.~Lett.~{\bf B385} (1996) 348; \\
S. Jadach, M. Skrzypek,
B. Pietrzyk, LAPP-EXP-99-01. }

\bibitem{bhwide}{S.~Jadach, W.~P\l{}aczek and B.F.L.~Ward,  
Phys.~Lett.~{\bf B390} (1997) 298; \\
W.~P\l{}aczek, S.~Jadach, M.~Melles,  B.F.L.~Ward and S.A.~Yost, 
CERN-TH/99-07, UTHEP-98-01,  {\tt hep-ph/9903381}.  }

\end{thebibliography}
\end{document}